\documentclass[aps,prb,citeautoscript,twocolumn,showpacs,preprintnumbers]{revtex4-1} 

\usepackage{graphicx}
\usepackage{amsfonts,amssymb,amsmath,amsthm}
\usepackage{hyperref}
\usepackage{tikz}

\newcommand{\ZZ}{\mathbb{Z}}
\newcommand{\nn}{\nonumber}
\newcommand{\omg}{\omega}
\newcommand{\mdag}{^\dagger}

\newcommand{\hc}{\text{H.c.}}

\newcommand{\id}{\openone}
\newcommand{\La}{\Lambda}
\newcommand{\T}{\mathcal{T}}
\newcommand{\ket}[1]{\left | #1 \right\rangle}
\newcommand{\bra}[1]{\left \langle #1 \right |}

\newtheorem{theorem}{Theorem}

\begin{document}

\title{Quantum Computing with Parafermions}
\author{Adrian Hutter}
\author{Daniel Loss}
\affiliation{Department of Physics, University of Basel, Klingelbergstrasse 82, CH-4056 Basel, Switzerland}

\date{\today}

\begin{abstract}
$\mathbb{Z}_d$ Parafermions are exotic non-Abelian quasiparticles generalizing Majorana fermions, which correspond to the case $d=2$.
In contrast to Majorana fermions, braiding of parafermions with $d>2$ allows to perform an entangling gate.
This has spurred interest in parafermions and a variety of condensed matter systems have been proposed as potential hosts for them.
In this work, we study the computational power of braiding parafermions more systematically.
We make no assumptions on the underlying physical model but derive all our results from the algebraical relations that define parafermions.
We find a familiy of $2d$ representations of the braid group that are compatible with these relations. The braiding operators derived this way reproduce those derived previously from physical grounds as special cases.
We show that if a $d$-level qudit is encoded in the fusion space of four parafermions, braiding of these four parafermions allows to generate the entire single-qudit Clifford group (up to phases), for any $d$.
If $d$ is odd, then we show that in fact the entire many-qudit Clifford group can be generated.
\end{abstract}

\maketitle

\section{Introduction}

Quasi-particles that live in two-dimensional space are described by the abstract theory of anyons \cite{wilczek,preskill}.
Of particular interest are non-Abelian anyons, those whose exchange statistics are non-commutative.
These exchange statistics only depend on homological properties of the trajectory of a particle, but are insensitive to small deformations.
This lead to the idea of using them to perform \emph{topological} quantum computations \cite{kitaev,nayak_rev,pachos}.

Quasi-particle modes in one- or two-dimensional condensed matter systems can carry (projective) non-Abelian statistics, too, allowing to identify them with non-Abelian anyon models.
Most prominently, the braiding statistics of localized Majorana zero modes are described by the Ising anyon model.
Parafermion modes are generalizations of Majorana fermions whose braiding behavior is more complex. 
The interest that parafermions have attracted in the condensed matter community in recent years \cite{fendley,lindner,cheng,you2012,clarke,vaezi,burrello,you,barkeshli_theory,mong,barkeshli1,klinovaja,zhang,oreg,klinovaja2,klinovaja3,barkeshli2,orth,alicea,Hutter15,klinovaja4} is due in part to the fact that they are computationally more powerful than Majorana fermions -- they allow to perform an entangling gate through quasi-particle braiding \cite{clarke}.

Proposals to physically realize parafermions typically require strong electron-electron interactions and thus often invoke edge states of fractional quantum Hall systems.
In this work, we are completely agnostic about the underlying physical system and derive all our results from the algebraic relations that define parafermions.
We want to study more systematically what quantum operations can be performed by braiding parafermions.
To this end, we first study what representations of the braid group are compatible with these algebraic relations.
For $\ZZ_d$ parafermions, we find a family of $2d$ representations.
The braiding behavior described by these $2d$ representations reproduces and generalizes that which has previously been derived from physical grounds for particular realizations of parafermions.

We then show that these $2d$ representations allow to generate the single-qudit Clifford group (up to global phases) through parafermion braiding (for any $d$), and the many-qudit Clifford group if $d$ is odd.
Finally, we briefly discuss extension of the Clifford group to universality.

Metaplectic anyons are a generalization of Majorana fermions that is different from, but related to parafermions \cite{metaplectic1}.
Their computational power has already been studied in-depth and some of them allow for universal quantum computation \cite{metaplectic1,metaplectic2,metaplectic3,metaplectic4}.

The rest of this work is ordered as follows.
In Sec.~\ref{sec:para} we introduce parafermions and their commutation relations, as well as the parity operator for a pair of parafermions.
Representations of the braid group that are compatible with these commutation relations are studied in Sec.~\ref{sec:rep},
and the phases that can be obtained by braiding of two parafermions under these representations are derived in Sec.~\ref{sec:braid}.
In Sec.~\ref{sec:log} we consider encoding a logical qudit into the fusion space of four parafermions and study the gates that can be performed on a single logical qudit.
Two-qudit entangling gates are studied in Sec.~\ref{sec:ent}. 
Finally, we discuss our results in Sec.~\ref{sec:disc} and briefly discuss extension of the Clifford group to universal quantum computing.

\section{Parafermion and parity operators}\label{sec:para}

For a totally ordered set $\lbrace i\rbrace$, $\ZZ_d$ parafermions are defined through the relations
\begin{align}\label{eq:parafermion}
 (\gamma_j)^d=\id\,, \quad \gamma_j\gamma_k = \omega^{\text{sgn}(k-j)}\gamma_k\gamma_j\,,
\end{align}
with $\omega=e^{2\pi i/d}$ (the dependence of $\omg$ on $d$ will be implicit henceforth).

The non-local commutation relations of parafermions can be obtained from the local commutation relations of $d$-dimensional generalizations of the Pauli matrices $X$ and $Z$ and a non-local transformation.
Indeed, let $X$ and $Z$ be defined over the relations $X^d = Z^d = \id$ and $ZX = \omega XZ$, and let $X_i$ and $Z_i$ denote the corresponding operators acting on the $i$-th of $n$ $d$-dimensional qudits. 
Then, the operators of $2n$ parafermions can be obtained from the Jordan-Wigner-like transformation \cite{fradkin,fendley}
\begin{align}\label{eq:JW}
 \gamma_{2i-1} = (\prod_{j<i}X_j)Z_i\,, \quad \gamma_{2i} = \omega^{(d+1)/2}(\prod_{j\leq i}X_j)Z_i\,.
\end{align}

In order to assess the potential of parafermions for topological quantum computing, we need to find the set of unitaries that can be performed by braiding them.
That is, for $2n$ parafermions we need to find a unitary representation of the braid-group $B_{2n}$. The braid group
\begin{align}
 B_{2n} = \left\langle \sigma_1, \sigma_2, \ldots, \sigma_{2n-1} \right\rangle
\end{align}
is generated by the $2n-1$ counterclockwise transpositions $\sigma_i$.
Here, $\sigma_i$ exchanges elements $i$ and $i+1$.
These satisfy ``far-commutativity''
\begin{align}
 \sigma_i\sigma_j = \sigma_j\sigma_i \qquad \text{if } |i-j|>1
\end{align}
and the Yang-Baxter equation
\begin{align}
 \sigma_i\sigma_j\sigma_i = \sigma_j\sigma_i\sigma_j \qquad \text{if } |i-j|=1\,.
\end{align}

Note that $(\gamma_i)^k(\gamma_{i+1})^l$ commutes with $\gamma_j\notin\lbrace\gamma_i, \gamma_{i+1}\rbrace$ iff $k+l\equiv0\text{ (mod }d)$.
Let us thus define the parity operators 
\begin{align}\label{eq:parity}
\La_i= \omg^{(d+1)/2}\gamma_i\gamma_{i+1}\mdag\,.
\end{align}
The prefactor ensures that $(\La_i)^d=\id$ for any $d$.
The parity operators satisfy
\begin{align}\label{eq:parityCommu}
 \La_i\La_j &= \La_j\La_i \qquad &\text{if } &|i-j|>1 \nn\\
 \La_i\La_j &= \omg^{\text{sgn}(j-i)}\La_j\La_i \qquad &\text{if } &|i-j|=1\,.
\end{align}
So the parity operators $\La_i$ are local operators. 
The phases of their eigenvalues can be interpreted in terms of local physical quantities, such as fractional charge, where the interpretation will depend on the specific model.

\section{Braid group representations}\label{sec:rep}

Let $U_i$ denote the unitary representation of $\sigma_i$, referred to as braid operator.
The unitary
\begin{align}\label{eq:ansatz}
 U_i = \frac{1}{\sqrt{d}}\sum_{m\in\ZZ_d}c_m(\La_i)^m
\end{align}
is the most general ansatz that commutes with all $\gamma_j\notin\lbrace\gamma_i, \gamma_{i+1}\rbrace$.
In particular, it guarantees that far-commutativity is satisfied.

Evidently, we have $\left[\La_i, U_i\right] = 0$. From Eqs.~(\ref{eq:parityCommu}) and (\ref{eq:ansatz}), it is also obvious that
\begin{align}
 \left[\La_i\La_{i+2}, U_{i+1}\right] = 0\,.
\end{align}
This means that for $2n$ parafermions  $\lbrace1,\ldots,2n\rbrace$, the overall parity $\La_1\La_3\ldots\La_{2n-1}$ is conserved by all braids $\lbrace U_1,\ldots,U_{2n-1}\rbrace$.

While far-commutativity is automatically satisfied by our ansatz, unitarity $U_iU_i\mdag=\id$ imposes the constraint
\begin{align}\label{eq:unitary}
 \forall r\in\ZZ_d : \sum_{m\in\ZZ_d}c_m\bar{c}_{m+r} = \delta_{r,0}d\,,
\end{align}
while the Yang-Baxter equation 
\begin{align}
U_iU_{i+1}U_i=U_{i+1}U_iU_{i+1} 
\end{align}
leads to the constraint
\begin{align}\label{eq:YB}
 \forall k,m\in\ZZ_d : \sum_{r\in\ZZ_d}c_rc_{k-r}c_m\omg^{mr} = \sum_{r\in\ZZ_d}c_rc_kc_{m-r}\omg^{kr}\,.
\end{align}
Similar equations have been derived in Ref.~\onlinecite{cobanera}.

We note that the mappings
\begin{align}\label{eq:trafo1}
 c_n \mapsto e^{i\phi}c_n\,
\end{align}
\begin{align}\label{eq:trafo2}
 c_n \mapsto \omg^nc_n\,
\end{align}
and 
\begin{align}\label{eq:trafo3}
 c_n \mapsto \bar{c}_{-n}
\end{align}
all leave conditions (\ref{eq:unitary}) and (\ref{eq:YB}) invariant.
Each solution to Eqs.~(\ref{eq:unitary}) and (\ref{eq:YB}) thus implies an entire family of solutions.
In order to fix the overall phase, we set $c_0\geq0$. 

\subsection{Small values of $d$}\label{sec:small}

In the following, we want to find solutions to Eqs.~(\ref{eq:unitary}) and (\ref{eq:YB}) for different values of $d$.
For $d=2$, it is easy to see that there are exactly two solutions ($c_0=1$, $c_1=\pm i$), leading to the well-known braiding operators $U_i$ for Majorana fermions \cite{nayak,ivanov}.
When $d=3$, we show in Appendix~\ref{app:Z3} that there are exactly $6$ solutions.
For $d=4$, we given an extensive set of solutions in Appendix~\ref{app:Z4}. This set is continuous, even when fixing the overall phase through the requirement $c_0\geq0$.
A discrete subset of these solutions was found in Ref.~\onlinecite{cobanera}.

\subsection{Arbitrary values of $d$}\label{sec:arbitrary}

A solution for arbitrary $d$ is given by 
\begin{align}\label{eq:general}
c_m = \omg^{m(m+d)/2}\,.
\end{align}
It is straightforward to verify Eqs.~(\ref{eq:unitary}) and (\ref{eq:YB}).
Note that the exponent can become half-integer if $d$ is even.
We also note that this solution satisfies $c_{m+d}=c_m$, so using elements of $\ZZ_d$ as indices is unproblematic.

The transformations Eq.~(\ref{eq:trafo2}) and Eq.~(\ref{eq:trafo3}) map the solution in Eq.~(\ref{eq:general}) to the $2d$ solutions
\begin{align}\label{eq:rep}
 c_m = \omg^{\pm m(m+2r+d)/2}\,,
\end{align}
with $r\in\ZZ_d$. 
The $2d$ solutions are specified through different choices of $r$ and $\pm$. 
By comparing the values of $c_1$ and $c_2$, one can easily show that these $2d$ solutions are indeed all distinct, except for $d=2$, where there are only the two solutions discussed in Sec.~\ref{sec:small}.
For $d=3$, these $2d$ solutions are all possible solutions, while for $d=4$, they form a discrete subset of a continuous set of solutions. 

For the rest of this work, we choose the sign $\pm$ in Eq.~(\ref{eq:rep}) to be $+$.
From Eq.~(\ref{eq:ansatz}), we see that choosing a different sign merely corresponds to exchanging $U_i$ and $U_i\mdag$, i.e., exchanging clockwise and anti-clockwise braids.
When studying how the braiding operator $U_i$ in Eq.~(\ref{eq:ansatz}) with the solution in Eq.~(\ref{eq:rep}) acts on the parafermion operators $\gamma_i$ and $\gamma_{i+1}$ by conjugation (it commutes with all other parafermion operators), 
we find the simple transformation law
\begin{align}\label{eq:trafo}
 \gamma_i &\mapsto \omg^{-r}\gamma_{i+1} \nn\\
 \gamma_{i+1} &\mapsto \omg^{1-r}\gamma_i\mdag(\gamma_{i+1})^2\,.
\end{align}
For even $d$, this transformation law was previously derived from physical grounds in Refs.~\onlinecite{clarke,vaezi}.

\section{Braiding operators}\label{sec:braid}

From Eqs.~(\ref{eq:JW}) and (\ref{eq:parity}) we find
\begin{align}
\La_{2i-1} = X_i\mdag\,,\qquad \La_{2i}=Z_iZ_{i+1}\mdag\,.
\end{align}
Since the spectrum of the generalized Pauli operators is given by $\lbrace1,\omg,\ldots,\omg^{d-1}\rbrace$, so is the spectrum of the parity operators.
For the parity operator $\La_i$, we can thus find an eigenbasis $\lbrace\ket{0}_i,\ket{1}_i,\dots,\ket{d-1}_i\rbrace$, where
\begin{align}\label{eq:basisDefi}
 \La_i\ket{m}_i=\omg^m\ket{m}_i\,.
\end{align}

Since the braiding operator $U_i$ commutes with $\La_i$, it is also diagonal in this basis. 
We find
\begin{align}\label{eq:UisDFT}
  U_i &= \frac{1}{\sqrt{d}}\sum_{m\in\ZZ_d}c_m(\La_i)^m \nn\\ &= \sum_{k\in\ZZ_d}\left(\frac{1}{\sqrt{d}}\sum_{m\in\ZZ_d}c_m\omg^{km}\right)\ket{k}\bra{k}_i \nn\\ 
 &= \sum_{k\in\ZZ_d}\check{c}_k\ket{k}\bra{k}_i\,.
\end{align}
So the phases we obtain through counterclockwise braiding of the two parafermions are given by the inverse discrete Fourier transform (DFT) of the coefficients $c_m$.
We have introduced the notation $\check{c}_k=\frac{1}{\sqrt{d}}\sum_{m\in\ZZ_d}c_m\omg^{km}$ to denote the inverse DFT, while we use $\hat{c}_k=\frac{1}{\sqrt{d}}\sum_{m\in\ZZ_d}c_m\omg^{-km}$ to denote the DFT.

The sequences $c_m$ in Eq.~(\ref{eq:rep}) are so-called Frank-Zadoff-Chu (FZC) sequences \cite{Frank63,Chu72,Beyme09}, used in modern cellular mobile communication systems for their favorable autocorrelation properties.
They are preserved under the DFT up to a prefactor and taking the complex conjugate. 
Indeed, we find  
\begin{align}\label{eq:relation}
 \check{c}_k 
 &= \frac{1}{\sqrt{d}}\sum_{m\in\ZZ_d}\omg^{m(m+2k+2r+d)/2} \nn\\
 &= \frac{1}{\sqrt{d}}\sum_{m\in\ZZ_d}\omg^{(m-k)(m+k+2r+d)/2} \nn\\
 &= \omg^{-k(k+2r+d)/2}\frac{1}{\sqrt{d}}\sum_{m\in\ZZ_d}\omg^{m(m+2r+d)/2} \nn\\
 &= \bar{c}_k\check{c}_0\,,
\end{align}
where $\check{c}_0=\omg^{-r(r+d)/2+d(1-d)/8}$.

The braid operator $U_i$ is thus given by
\begin{align}\label{eq:UU}
 U_i &= \check{c}_0\sum_{k\in\ZZ_d}\bar{c}_k\ket{k}\bra{k}_i \nn\\
&= \check{c}_0\sum_{k\in\ZZ_d}\omg^{- k(k+2r+d)/2}\ket{k}\bra{k}_i \nn\\
&\propto \sum_{k\in\ZZ_d}\omg^{- (k+r+d/2)^2/2}\ket{k}\bra{k}_i\,.
\end{align}
Braiding operators of this form have been derived from different physical models \cite{lindner,clarke,vaezi,you,Hutter15}.

\section{Logical Qudits}\label{sec:log}

The $d$ states $\lbrace\ket{k}_i\rbrace$ defined through Eq.~(\ref{eq:basisDefi}) form a basis of the fusion space of the two parafermions $\gamma_i$ and $\gamma_{i+1}$, corresponding to different eigenvalues of their parity operator $\La_i$. 
A $d$-dimensional qudit is thus naturally associated with each pair of parafermions.
Each parafermion asymptotically adds a factor $\sqrt{d}$ to the groundstate degeneracy, which by definition gives its quantum dimension.

However, powers of the the braid operator $U_i$ given in Eq.~(\ref{eq:UU}) are the only gates that can be performed locally on this qudit.
In particular, it is not possible to evolve from a state $\ket{k}_i$ to $\ket{l}_i$ with $k\neq l$.
Following the standard procedure for Majorana fermions \cite{pachos,bravyi}, we thus encode one qudit into the fusion space of \emph{four} parafermions.

For concreteness, let us consider the set of parafermions $\lbrace1,2,3,4\rbrace$.
Their joint fusion space has dimension $d^2$. For computational purposes, we restrict to the $d$-dimensional subspace for which $\La_1\La_3=1$, i.e.\ the states of the four parafermions with neutral parity. We can act on this space by the group of unitaries generated by $U_1$, $U_2$, and $U_3$.
Recall that all of these commute with $\La_1\La_3$ and hence preserve the computational subspace.

The computational subspace of states with $\La_1\La_3=1$ is spanned by the logical states $\lbrace\ket{0}_L,\ldots,\ket{d-1}_L\rbrace$, where $\ket{k}_L=\ket{k}_1\otimes\ket{d-k}_3$.
We can introduce generalized Pauli operators $X$ and $Z$ acting on this subspace which are defined over their matrix elements
\begin{align}
 &\bra{k}_LX\ket{l}_L = \delta_{k,l\oplus1} \nn\\ &\bra{k}_LZ\ket{l}_L = \omg^k\delta_{k,l}\,.
\end{align}
(We denote addition modulo $d$ by $\oplus$.)
Let us introduce the superoperator $\T$ to denote restriction to the computational subspace.
Then we have $\T(\La_1)=Z$, $\T(\La_2)=X$, and $\T(\La_3)=Z\mdag$.

The operators $\T(U_1)$ and $\T(U_3)$ are both diagonal in the computational basis with
\begin{align}
 \bra{k}_LU_1\ket{k}_L = \bra{k}_1U_1\ket{k}_1 = \check{c}_k
\end{align}
and
\begin{align}
 \bra{k}_LU_3\ket{k}_L = \bra{d-k}_3U_3\ket{d-k}_3 = \check{c}_{-k} = \hat{c}_k
\end{align}
(recall that $\hat{\hat{c}}_m=c_{-m}$).
However, $\T(U_2)$ is not diagonal.
The operators $\T(U_1)$ and $\T(U_2)$ are isospectral and diagonal in the eigenbasis of $Z$ and $X$, respectively. 
These eigenbases are related by a DFT. Formally, let
\begin{align}
 F = \frac{1}{\sqrt{d}}\sum_{k,m\in\ZZ_d}\omg^{km}\ket{k}\bra{l}_L\,.
\end{align}
The DFT $F$ generalizes the qubit Hadamard gate and satisfies $FXF\mdag=Z$ and $FZF\mdag=X\mdag$ .
Thus,
\begin{align}
 \bra{k}_LU_2\ket{l}_L
 &= \bra{k}_LF\mdag U_1F\ket{l}_L \nn\\
 &= \sum_{r\in\ZZ_d}\check{c}_r\bra{k}_LF\mdag\ket{r}_L\bra{r}_LF\ket{l}_L \nn\\
 &= \frac{1}{d}\sum_{r\in\ZZ_d}\check{c}_r\omg^{-r(k-l)} \nn\\
 &= \frac{1}{\sqrt{d}}c_{k-l}\,.
\end{align}

The gates $\T(U_1)$ and $\T(U_1U_2U_1)$ generate all operations that can be performed on the logical subspace of a single logical qudit through braiding of the four parafermions.
Let us study how they act on the generalized Pauli operators $X$ and $Z$ by conjugation.

Recall from Eqs.~(\ref{eq:UisDFT}) and (\ref{eq:relation}) that
\begin{align}
 U_i=\check{c}_0\sum_{k\in\ZZ_d}\bar{c}_k\ket{k}\bra{k}_i
\end{align}
and hence 
\begin{align}
  \T(U_1)=\check{c}_0\sum_{k\in\ZZ_d}\bar{c}_k\ket{k}\bra{k}_L\,.
\end{align}
Clearly, $\T(U_1)$ and $Z$ commute. 
For the action of $\T(U_1)$ on $X$, we find, making use of our general solution Eq.~(\ref{eq:rep}),
\begin{align}
 \T(U_1)X\T(U_1)\mdag 
 &= \sum_{k\in\ZZ_d}\ket{k\oplus1}\bra{k}_L\bar{c}_{k+1}c_k \nn\\
 &= \sum_{k\in\ZZ_d}\ket{k\oplus1}\bra{k}_L\omg^{-(2r+d+1)/2}\omg^{- k} \nn\\
 &= \omg^{-(2r+d+1)/2}XZ\mdag\,.
\end{align}

Now let us compute
\begin{align}
 \bra{k}_LU_1U_2U_1\ket{l}_L 
 &= \check{c}_k\bra{k}_LU_2\ket{l}_L\check{c}_l \nn\\
 &= \frac{1}{\sqrt{d}}\check{c}_kc_{k-l}\check{c}_l\,.
\end{align}
At this point, we make use of Eq.~(\ref{eq:relation}) to find
\begin{align}
  \bra{k}_LU_1U_2U_1\ket{l}_L 
 = \frac{1}{\sqrt{d}}(\check{c}_0)^2\bar{c}_kc_{k-l}\bar{c}_l.
\end{align}
After inserting the general form of our $2d$ solutions, Eq.~(\ref{eq:rep}), we finally arrive at
\begin{align}
 \bra{k}_LU_1U_2U_1\ket{l}_L = (\check{c}_0)^2\frac{1}{\sqrt{d}}\omg^{kl}\,.
\end{align}
We can thus make the identification
\begin{align}
 \T(U_1U_2U_1) = (\check{c}_0)^2F\,.
\end{align}

To summarize, the action of the braiding operator $U_1$ on the logical operators $X$ and $Z$ by conjugation is given by
\begin{align}\label{eq:phase}
 X &\mapsto XZ\mdag \nn\\
 Z &\mapsto Z\,,
\end{align}
up to phases, while the action of $U_1U_2U_1$ is given by
\begin{align}\label{eq:hadamard}
 X &\mapsto Z \nn\\
 Z &\mapsto X\mdag\,.
\end{align}

These braids thus map products of Pauli operators to other products of Pauli operators under conjugation. 
Such gates are known as Clifford gates, and the Clifford group on a given number of qudits is defined as the group of unitaries that map any product of Pauli operators on these qudits to another such product.
Farinholt has shown that the transformations in Eqs.~(\ref{eq:phase}) and (\ref{eq:hadamard}) are a necessary and sufficient set of gates for generating the single-qudit Clifford group (up to phases), for arbitrary $d$ \cite{farinholt}.
We can thus conclude the following.

\begin{theorem}
 Braiding of $\ZZ_d$ parafermions allows one to generate the entire single-qudit Clifford group (up to phases), for any qudit dimension $d$.
\end{theorem}

This theorem applies to all $2d$ representations of the braid group given by Eq.~(\ref{eq:rep}).

\section{Entangling gates}\label{sec:ent}

\begin{figure}
\begin{tikzpicture}[xscale=1.00, yscale=1.00]
\node at (2,0.5) {3};
\node at (3,0.5) {4};
\draw [dashed] (3.5,0.3) -- (3.5,0.8);
\node at (4,0.5) {5};
\node at (5,0.5) {6};
\node at (6,0.5) {7};
\draw (2,1) -- (2,2);
\draw (3,1) -- (3.4,1.4);	\draw (3.6,1.6) -- (4,2);
\draw (2,2) -- (2.4,2.4);	\draw (2.6,2.6) -- (3,3);
\draw (4,1) -- (2,3);
\draw (5,1) -- (5,3);
\draw (6,1) -- (6,4);
\draw (4,2) -- (4,3);
\draw (5,3) -- (4.6,3.4);	\draw (4.4,3.6) -- (4,4);
\draw (4,3) -- (6,5);
\draw (6,4) -- (5.6,4.4);	\draw (5.4,4.6) -- (5,5);
\draw (4,4) -- (4,5);
\draw (4,5) -- (3.6,5.4);	\draw (3.4,5.6) -- (3,6);
\draw (3,3) -- (3,5);
\draw (3,5) -- (5,7);
\draw (5,5) -- (5,6);
\draw (5,6) -- (4.6,6.4);	\draw (4.4,6.6) -- (4,7);
\draw (5,7) -- (4.6,7.4);	\draw (4.4,7.6) -- (4,8);
\draw (4,7) -- (6,9);
\draw (6,5) -- (6,8);
\draw (4,8) -- (4,9);
\draw (3,6) -- (3,9);
\draw (6,8) -- (5.6,8.4);	\draw (5.4,8.6) -- (5,9);
\draw (4,9) -- (3.6,9.4);	\draw (3.4,9.6) -- (3,10);
\draw (3,9) -- (5,11);
\draw (5,9) -- (5,10);
\draw (5,10) -- (4.6,10.4);	\draw (4.4,10.6) -- (4,11);
\draw (3,10) -- (3,11);
\draw (4,11) -- (4,12);
\draw (3,11) -- (2.6,11.4);	\draw (2.4,11.6) -- (2,12);
\draw (2,3) -- (2,11);
\draw (2,11) -- (4,13);
\draw (5,11) -- (5,13);
\draw (6,9) -- (6,13);
\draw (4,12) -- (3.6,12.4);	\draw (3.4,12.6) -- (3,13);
\draw (2,12) -- (2,13);
\end{tikzpicture}
\caption{Illustration of the braid $S$ which acts like $(C_X)^{-2}$ on the computational subspace. Time flows upwards. The small dashed line shows the separation between logical qudits $A$ and $B$.}
\label{fig:braid} 
\end{figure}
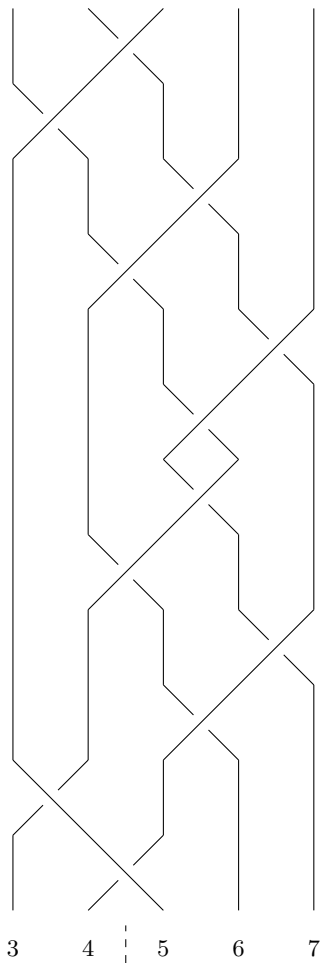

Now let us consider two qudits encoded in the parafermion quadruplets $A=\lbrace1,2,3,4\rbrace$ and $B=\lbrace5,6,7,8\rbrace$, respectively.
Consider the braids $V=U_4U_3$, $W=U_5U_4U_6U_5$, and $S=VW^2V\mdag$.
An illustration of this braid is given in Fig.~\ref{fig:braid}. 

It is a tedious but straightforward task to show that $S$ acts as follows by conjugation,
\begin{align}
 \La_1 &\mapsto \La_1 \nn\\
 \La_2 &\mapsto \La_2(\La_6)^{-2} \nn\\    
 \La_3 &\mapsto \La_3 \nn\\
 \La_5 &\mapsto (\La_3)^{-2}\La_5 \nn\\
 \La_6 &\mapsto \La_6 \nn\\
 \La_7 &\mapsto (\La_3)^{2}\La_7\,.
\end{align}
In particular, it follows that $S\La_1\La_3S\mdag=\La_1\La_3$ and $S\La_5\La_7S\mdag=\La_5\La_7$, and hence that $S$ preserves the computational subspace.

Recall that  
\begin{align}
&\T(\La_1)=Z_A \quad \T(\La_2)=X_A \quad \T(\La_3)=Z_A\mdag \quad \nn\\
&\T(\La_5)=Z_B \quad \T(\La_6)=X_B \quad \T(\La_7)=Z_B\mdag\,.
\end{align}
The braid $S$ thus acts as follows on the computational subspace,
\begin{align}\label{eq:entangling}
Z_A &\mapsto Z_A \nn\\
X_A &\mapsto X_A X_B^{-2}\nn\\
Z_B &\mapsto Z_A^2 Z_B \nn\\
X_B &\mapsto X_B\,.
\end{align}

For two $d$-dimensional qudits $A$ and $B$, the entangling gate $C_X$ is defined over
\begin{align}
 C_X\ket{i}_A\ket{j}_B = \ket{i}_A\ket{i\oplus j}_B\,.
\end{align}
It acts by conjugation as 
\begin{align}
Z_A &\mapsto Z_A \nn\\
X_A &\mapsto X_A X_B\nn\\
Z_B &\mapsto Z_A\mdag Z_B \nn\\
X_B &\mapsto X_B\,.
\end{align}
We can thus make the identification
\begin{align}
\T(S\mdag) = (C_X)^2\,.
\end{align}

Similarly, one can show that the simpler braid $T=(U_4U_3U_5U_4)^2$ performs the square of the controlled phase-gate on the computational subspace.
That is, $\T(T) = (C_Z)^2$, where $C_Z\ket{i}_A\ket{j}_B=\omg^{ij}\ket{i}_A\ket{j}_B$.

Clearly, $(C_X)^d=\id$. If $d$ is odd, we thus have
\begin{align}
\T(S^{-(d+1)/2}) = C_X\,.
\end{align}
Applying the braid $S\mdag$ $\frac{d+1}{2}$ times thus allows one to apply the gate $C_X$ to the computational subspace of two logical qudits.
Together with the single-qudit gates that transform the logical operators $X$ and $Z$ as in Eqs.~(\ref{eq:phase}) and (\ref{eq:hadamard}), the entangling gate $C_X$ generates the entire many-qudit Clifford group \cite{farinholt}.
This allows us to conclude the following.

\begin{theorem}
If $d$ is odd, braiding of $\ZZ_d$ parafermions allows one to generate the entire many-qudit Clifford group (up to phases).
\end{theorem}

\section{Discussion}\label{sec:disc}

Here, we have shown that if $d$ is odd, then braiding of $\ZZ_d$ parafermions allows to generate the full Clifford group, up to phases.
While even values of $d>2$ allow to perform non-trivial entangling gates, we were not able to show the same for these parafermions, and suspect that only a non-trivial subset of the Clifford group can be generated.
The Clifford group by itself is not universal for quantum computing. 
In fact, the operations that can be obtained from Clifford unitaries and preparation and measurement in the computational basis are known as stabilizer operations, and these can be efficiently simulated on a classical computer \cite{gottesman_phd,gottesman}.

However, if $d$ is prime, it is known that Clifford unitaries together with an arbitrary non-Clifford gate are sufficient for universal quantum computing \cite[Appendix D]{campbell_magic}.
Letting two parafermion modes interact with each other (e.g.\ by bringing them close to each other) will allow to perform a unitary that generically is not a Clifford gate. 
A natural interaction between parafermions $1$ and $2$ such as $\alpha\gamma_1\gamma_2\mdag+\hc$ will commute with the parity operator $\La_1$ and thus preserve the computational subspace.
Indeed, the computational subspace is preserved as long as the quantity measured by the phases of the parity operator is conserved.

A non-Clifford gate that is performed by non-topological means will not be fault-tolerant.
Fault-tolerance can be restored using only stabilizer operations through magic state distillation \cite{bravyi_magic}, which has been studied in-depth for prime-dimensional qudits \cite{campbell_magic,campbell, campbell_qutrit}.

Given our results and the work on magic state distillation, $\ZZ_d$ parafermions where $d$ is an odd prime seem most attractive from a quantum information processing perspective.
Unfortunately, many proposals to physically realize parafermions concern the case where $d$ is even \cite{lindner,clarke,vaezi,burrello,orth,Hutter15}. 
On the other hand, $\ZZ_3$ parafermions can for example emerge in interacting nanowires \cite{oreg,klinovaja,klinovaja2}.

Finally, we note that that when $\ZZ_d$ parafermions arise as ends of defect lines in $D(\ZZ_d)$ quantum double models \cite{you2012,you}, the entire Clifford group can be generated through quasi-particle braiding for any $d$ by making use of the Abelian excitations of the underlying state. This includes the case $d=2$, i.e., Majorana fermions in a qubit toric code. 
Ref.~\onlinecite{Hutter15} describes in detail how this can be achieved for $d=4$, and the generalization to arbitrary $d$ is straightforward.

\section{Acknowledgements}

We thank Jelena Klinovaja and James R.~Wootton for helpful discussions.
This work was supported by  the Swiss NF and NCCR QSIT.


\newpage

\appendix

\section{Braiding operators}

\subsection{Solutions for $d=3$}\label{app:Z3}
From Eqs.~(\ref{eq:unitary}) and (\ref{eq:YB}), we get the following set of non-equivalent equations,
\begin{align}\label{eq:alleqs}
 &|c_0|^2+|c_1|^2+|c_2|^2 = 3 \nn\\
 &c_0\bar{c}_1+c_1\bar{c}_2+c_2\bar{c}_0 = 0 \nn\\
 &c_0\bar{c}_2+c_1\bar{c}_0+c_2\bar{c}_1 = 0 \nn\\
 &(c_0)^2c_1 + (c_1)^2c_2 + (c_2)^2c_0 = 0 \nn\\
 &(c_0)^2c_2 + (c_1)^2c_0 + (c_2)^2c_1 = 0 \nn\\
 &(c_1)^3 = (c_2)^3\,.
\end{align}
In order to satisfy the last equation, we set $c_2=c_1\omg^r$ with $r\in\ZZ_3$.
With this identification, the third and fourth equation become equivalent and we are left with three equations,
\begin{align}
 &|c_0|^2+2|c_1|^2 = 3 \nn\\
 &c_0\bar{c}_1+c_1\bar{c}_1\omg^{-r}+c_1\bar{c}_0\omg^r = 0 \nn\\
 &(c_0)^2c_1 + (c_1)^3\omg^r + (c_1)^2c_0\omg^{-r} = 0 \,.
\end{align}
Defining $c_1=se^{i\phi}$, and recalling that we assume $c_0\geq0$, we find the three equations
\begin{align}\label{eq:system}
 &(c_0)^2+2s^2 = 3 \nn\\
 &c_0(\omg^re^{-i\phi}+\omg^{-r}e^{i\phi})+s = 0 \nn\\
 &c_0se^{i\phi}+(c_0)^2\omg^r+s^2e^{i2\phi}\omg^{-r} = 0 
\end{align}
for the three \emph{real} unknowns $c_0$, $s$, and $\phi$.
Solving the last equation for $e^{i\phi}$, we find
\begin{align}
 e^{i\phi} = \frac{c_0}{s}\omg^re^{\pm i 2\pi/3}\,.
\end{align}
Since $c_0$ and $s$ are non-negative, we conclude that $c_0=s=1$ and are left with a single non-trivial equation
\begin{align}
 e^{i\phi} = \omg^{r\pm1}\,.
\end{align}
The three possible values $r\in\ZZ_3$ lead to the following six solutions,
\begin{align}
 c_0 = 1 \qquad & c_1 = 1 \qquad c_2 = \omg \nn\\
 c_0 = 1 \qquad & c_1 = \bar{\omg} \qquad c_2 = \bar{\omg} \nn\\
 c_0 = 1 \qquad & c_1 = \omg \qquad c_2 = 1 \nn\\
 c_0 = 1 \qquad & c_1 = \bar{\omg} \qquad c_2 = 1\nn\\
 c_0 = 1 \qquad & c_1 = 1 \qquad c_2 = \bar{\omg} \nn\\
 c_0 = 1 \qquad & c_1 = \omg \qquad c_2 = \omg \,.
\end{align}
Note that the first and the last three solutions are related to each other through transformation Eq.~(\ref{eq:trafo2}), while the two sets are related to each other trough transformation Eq.~(\ref{eq:trafo3}).

\subsection{Solutions for $d=4$}\label{app:Z4}

For $d=4$, we get from Eqs.~(\ref{eq:unitary}) and (\ref{eq:YB}) the following non-equivalent equations:
\begin{align}\label{eq:eqs4}
 &|c_0|^2+|c_1|^2+|c_2|^2+|c_3|^2=4 \nn\\
 &c_0\bar{c}_1+c_1\bar{c}_2+c_2\bar{c}_3+c_3\bar{c}_0=0 \nn\\
  &c_0\bar{c}_2+c_1\bar{c}_3+c_2\bar{c}_0+c_3\bar{c}_1=0 \nn\\
 &c_1((c_0)^2+(c_2)^2)+2c_0c_2c_3=0 \nn\\
 &c_3((c_0)^2+(c_2)^2)+2c_0c_1c_2=0 \nn\\
 &c_0((c_1)^2+(c_3)^2)+(c_0)^2c_2-(c_2)^3+2c_1c_2c_3=0 \nn\\
 &(c_1)^2=(c_3)^2 .
\end{align}
From the last line, we have $c_1=\pm c_3$. For $c_1=c_3$, we straightforwardly find $c_0=-c_2$, and for $c_1=-c_3$, we find $c_0=c_2$. In either case, we find $|c_0|=|c_1|=|c_2|=|c_3|$.
Up to an overall phase, an extensive set of solutions is thus given by
\begin{align}\label{eq:allsols4}
 c_0 = 1\,,\qquad c_1=e^{i\phi}\,,\qquad c_2=\pm1\,,\qquad c_3=\mp e^{i\phi}\,,
\end{align}
where $\phi\in[0,2\pi)$ is arbitrary.
We note that in contrast to the case $d=3$, the set of solutions is continuous for $d=4$ (even when ignoring an overall phase).


\end{document}